\documentclass{article}

\usepackage{amssymb,stmaryrd}

\begin{document}

\title{The physical Church-Turing thesis and 
non deterministic
computation over the real numbers} 
\author{Gilles Dowek\thanks{INRIA, 23 avenue d'Italie, CS 81321, 
75214 Paris Cedex 13, France, \texttt{gilles.dowek@inria.fr},
\texttt{http://www-roc.inria.fr/who/Gilles.Dowek/}.}}  \date{}
\maketitle

\begin{abstract}
On the real numbers, the notions of a semi-decidable relation and that
of an effectively enumerable relation differ.  The second only seems
to be adequate to express, in an algorithmic way, non deterministic
physical theories, where magnitudes are represented by real numbers.
\end{abstract}

{\small Keywords: physical Church-Turing thesis, computability over
the real numbers, non determinism}

\bigskip

\thispagestyle{empty}

The physical Church-Turing thesis, if it holds, suggests that the laws
of nature can be expressed, not only in the language of mathematics,
but also in an algorithmic language.  Gandy's argument, that support
the physical Church-Turing thesis, even suggests that these algorithms
may operate on the elements of a countable set, such as the natural
numbers, the rational numbers, the elements of an extension of finite
degree of the field of rational numbers, the computable real numbers,
{\em etc.}  

Nevertheless, the desire to remain as close as possible to
the traditional formulations of the laws of nature rather leads to try
to describe these laws using algorithms operating on real numbers.
This
use of algorithms operating on real numbers may be just a first step
towards that of algorithms operating on the elements of a countable set, but
it may also simplify the formulation of the laws of nature, that might be more
complex, if formulated as algorithms operating on the elements of a
countable set.
We do not claim that physical magnitudes are adequately described by
real numbers, neither that all the infinite amount of information
contained in a real number is relevant in this description, but simply
that real numbers have been used to describe physical magnitudes and
that this justifies the attempt to describe the laws of nature in an
algorithmic way, using algorithms operating on real numbers.

An algorithmic formulation of the laws of nature should not predict
more than the usual propositional formulation of the same laws. As
some of the laws of physics are non deterministic, it seems that we
must use non deterministic algorithms to express these laws in an
algorithmic way.

The notion of a non deterministic algorithm and that of an algorithm
operating on real numbers have been defined and well studied. It seems
however that the conjunction of these two notions, {\em i.e.} the notion
of {\em non deterministic algorithm operating on real numbers}, remains to
be defined. We discuss in this paper what such a definition could be.

We assume that the notion of a computable function operating on the
natural numbers is known. The set of such functions can, for instance,
be defined as the smallest set containing the projections, the null
functions, the successor function and closed by composition,
definitions by induction and minimization. Other definitions based on
the notions of Turing machines, rewrite systems, {\em etc.}
are possible. This notion extends to countable sets by
numbering their elements.

\section{Non deterministic algorithms}

\subsection{Algorithms, functions and relations}

A deterministic algorithm, that takes an element of a set $A$ as an
argument and returns an element of a set $B$ if it terminates,
defines a partial function from $A$ to $B$. 

A non deterministic algorithm, in contrast, defines a relation between
the sets $A$ and $B$ and it is therefore natural to try to
characterize the relations that correspond to these algorithms.  
A relation $R$ between two sets $A$ and $B$ can always be 
seen as a function from $A$ to the powerset of $B$: the function
that maps the element $x$ to the set $R_x = \{y \in B~|~x~R~y\}$. 
This leads to raise, in a first step,
the question of the representation of sets with computable functions.

\subsubsection{Characteristic functions}

When the set $B$ is countable and the sets $R_x$ are all finite, they
can be numbered.  But, this solution is not available when, as it is
often the case in physics, the sets $R_x$ may be infinite.

In this case, a possibility is to represent the set $R_x$ is by an algorithm
expressing its characteristic function. 
Using this representation of sets leads to represent the
relation $R$ by a function $f$ that maps $x$ to a function that maps
$y$ to the value $1$ or $0$ depending on the fact that $x~R~y$ or not.
Using the fact that a function taking values in a functional space is
equivalent to a function of several arguments, the relation $R$ can
also be represented as the binary function that maps $(x,y)$ to $1$ or
$0$ depending on the fact that $x~R~y$ or not, {\em i.e.} by its own
characteristic function.  
This leads to represent a non deterministic
algorithm by a {\em decidable relation}, {\em i.e.} a relation whose
characteristic function is computable.

\subsubsection{Partial characteristic functions}

An other possibility is to represent the set $R_x$ by 
an algorithm expressing its 
partial
characteristic function, {\em i.e.} by the function that takes the value
$1$ at $y$ if $y$ is an element of $R_x$, but is not defined at $y$
otherwise.  

Using this representation of sets leads to represent the relation $R$
by a function $f$ that maps $x$ to the partial function that maps $y$
to the value $1$ if $x~R~y$ and that is not defined at $y$ otherwise.
Using again the fact that a function taking values in a functional space is
equivalent to a function of several arguments, the relation $R$ can
also be represented as a binary function 
that maps $(x,y)$ to $1$ if $x~R~y$ and that is
not defined at $(x,y)$ otherwise, {\em i.e.} by its own partial
characteristic function.
This leads to represent a non deterministic
algorithm by a {\em semi-decidable relation}, {\em i.e.} a relation whose
partial characteristic function is computable.

\subsubsection{Enumeration functions}

A third possibility is to represent the set $R_x$ by an algorithm
expressing a total function whose image is this set.  The elements of
the domain $\Omega$ of this function are called {\em indices}. When
this set is finite or countable, by using a numbering of its the
elements, we can restrict, without loss of generality, to the case
where it is the set of natural numbers. To avoid to distinguish a
special case for the empty set, we shall consider that this function
may fail, {\em i.e.} take a special value that is not taken into
account in its image.

Using this representation of sets leads to represent the relation $R$
by a function $f$ that maps $x$ to a function $f_x$ such that $x~R~y$
if and only if there exists an index $i$ such that $y = f_x(i)$.
Without loss of generality, we can assume that all the functions $f_x$
share the same set of indices and using again the fact that a function taking
values in a functional space is equivalent to a function of several
arguments, the relation $R$ can also be represented by a binary
function $f$ such that $x~R~y$ if and only if there exists an index
$i$ such that $y = f(x,i)$.

This leads to represent non deterministic algorithms by
{\em effectively enumerable} relations, 
{\em i.e.} 
relations $R$ such that there exists
exists a set $\Omega$ and a computable function $f$ from $A \times
\Omega$ to $B$ such that for all $x$ and $y$, $x~R~y$ if and only if 
there exists an $i$ such that $y = f(x,i)$.

\subsubsection{Relating these notions}

These various notions are related: classical results show
that a decidable
relation is semi-decidable and that a relation is semi-decidable if
and only if it is effectively enumerable.

Indeed, if a relation $R$ is decidable, then by composing its characteristic
function with a function that maps $1$ to $1$ and is not defined at
$0$, we obtain its partial characteristic function, that is therefore
computable.

If a relation $R$ is semi-decidable, then there exists a computable
function $f$ that takes the value $1$ at $(x,y)$ if and only if
$x~R~y$. Then there exists a computable function $g$ such
that $g(x,y,i)$ takes the value $1$ if the $i$ first steps of the
computation of the value of $f$ at $(x,y)$ give the result $1$
and takes the value $0$ otherwise. Let $h$ be the computable function
mapping $(x,(y,i))$ to $y$ if $g(x,y,i) = 1$ and an exceptional value
otherwise. We have $x~R~y$ if and only if 
there exists a $j$ such that $y = h(x,j)$.
Thus the relation $R$ is effectively enumerable.

Remark that we can avoid the use of an exceptional value, 
if all the sets $R_x$ are non empty. In this case, we first show 
the existence of a computable function $d$ mapping
$x$ to an element of $R_x$ and we define the binary function $h$ as
the function that maps $(x,(y,i))$ to $y$ if $g(x,y,i) = 1$ and to
$d(x)$ otherwise.

Conversely, if a relation $R$ is effectively enumerable, there exists
a function $f$ such that $x~R~y$ if and only if there exits an index
$i$ such that $y = f(x,i)$. By numbering the elements of the set
$\Omega$, we can restrict, without loss of generality, to the case where
this set is the set of natural numbers. To determine if two elements
$x$ and $y$ are in relation or not, it is sufficient to compute, in a
sequential way, the values $f(x,0)$, $f(x,1)$, $f(x,2)$,
...  and to return the value $1$ if the result of one of these
computations is equal to $y$. The partial characteristic function of
$R$ is therefore computable. 

Remark that the decidability of equality over the natural numbers is
used in this proof in an essential way.

Finally, we can prove that a relation can be defined in a non
deterministic classical model (Turing machine, rewrite system, {\em etc.})
if and only if it is semi-decidable, of
equivalently effectively enumerable.  
Thus, the notion that captures non deterministic algorithms is
not that of a decidable relation, but that
that of a semi-decidable relation or,
equivalently, that of an effectively enumerable relation.

\subsection{Index, elementary event, possible world, universe and
hidden variable}

This notion of an index, used in the definition of the notion of an
effectively enumerable relation, is reminiscent of
several other notions, used in various domains, that are
all concerned, in a way or an other, with the notion of potentiality.

First, it is reminiscent of the notion of an {\em elementary event}
in Kolmogorov's notion of a probability space. 
Given an element $x$, the function 
that maps an index $i$ to the value $f(x,i)$ could therefore be called a {\em
random variable}, even if the set of indices is not necessarily
equipped with a probability distribution.  Then, it is reminiscent of
the notion of a {\em possible world} that is used in Kripke's models
of modal logics, and that of {\em universe} in Everett's
interpretation of quantum physics, according to which all the possible
results of an experiment, predicted by the theory, indeed happen, but
each in a different universe. The fact that $f(x,i) = y$ can therefore
be read as the fact that, in the world $i$, the algorithm $f$ maps
$x$ to $y$.  Finally, it is reminiscent of the notion of a {\em hidden
  variable}, because non-determinism is interpreted as the dependence
to an ancillary variable.

\section{Computation over the real numbers}

The links between computability theory and analysis are as old
as computability theory itself, as Turing already defined the notion of 
a computable real number. Here, we are interested in defining a notion of 
a computable function from the real numbers to the real numbers.

\subsection{Computability over the real numbers}

A possible definition, in the style 
Grzegorczyk, Lacombe, and 
Weihrauch is that a function $f$
from an interval $I$ of ${\mathbb R}$ to ${\mathbb R}$ is computable
if there exists a computable function $F$ from ${\mathbb Q} \times 
({\mathbb Q}^{+} \setminus \{0\})$ to 
${\mathbb Q} \times ({\mathbb Q}^{+} \setminus \{0\} \cup \{+ \infty\})$
such that for all $x$ in $I$, $q$ and $r$ in ${\mathbb Q}$, $\eta$ and
$\varepsilon$ in ${\mathbb Q}^{+} \setminus \{0\}$
$$(F(q,\eta) = (r,\varepsilon)~\mbox{and}~|x - q| \leq \eta)
\Rightarrow |f(x) - r| \leq \varepsilon$$ Put in another way, if
$(r,\varepsilon) = F(q,\eta)$ and $q$ is an approximation of $x$ up to
$\eta$, then $r$ is an approximation of $f(x)$ up to $\varepsilon$.

We must however be able to know the value of $f(x)$ with an arbitrary
accuracy, which leads to put another condition: if $x$ is a real
number and $(q_n)_n$, $(\eta_n)_n$, $(r_n)_n$ and $(\varepsilon_n)_n$
are sequences such that for all $n$, $(r_n,\varepsilon_n) =
F(q_n,\eta_n)$ and $|x - q_n| \leq \eta_n$ and the sequence $\eta$
goes to $0$ at infinity, then the sequence $\varepsilon$ also
goes to $0$ at infinity.

This definition is equivalent to that of Weihrauch, and 
when $I$ is a compact interval $[a,b]$, it is also equivalent to the 
definition of 
Grzegorczyk and Lacombe, as, for instance, presented by Pour El and Richards, 
where a function $f$ from an interval $I$ of
${\mathbb R}$ to ${\mathbb R}$ is computable if there exists a computable
function $e$ from ${\mathbb Q}^{+} \setminus \{0\}$ to 
${\mathbb Q}^{+} \setminus \{0\}$ and a
computable function $F$ from 
${\mathbb Q} \times ({\mathbb Q}^{+} \setminus \{0\})$ 
to ${\mathbb Q}$ such that for all $x$ in $I$, $q$ in ${\mathbb Q}$ and
$\varepsilon$ in ${\mathbb Q}^{+} \setminus \{0\}$
$$|x - q| \leq e(\varepsilon)  \Rightarrow |f(x) - F(q,\varepsilon)| \leq
\varepsilon$$ 
and where the function $e$ mapping the
desired accuracy $\varepsilon$ on the image of $f$ to the requested
accuracy on its argument, has to be known {\em a priori} and uniform
on the interval $I$.  
But these two definitions differ when the interval $I$ is not 
compact.

Remark that all computable functions are continuous by
definition. Hence the characteristic function of equality and of the set
${\mathbb R}^+ \setminus \{0\}$ are not computable.

We leave out of this discussion several other important notions of
computability over the real numbers. The first is the notion proposed
by Blum, Cucker, Shub, and Smale where equality is a decidable
relation.  Another is the non extensional notion of computability,
where the value of a function at a real number $x$ may depend not only
on this real number $x$, but also on its presentation, for instance on
the Cauchy sequence used to define it. Despite their intrinsic
interest, these notions do not seem to be relevant to the question of
the algorithmic description of the laws of nature discussed here.

Remark that, when extending the notion of a computable function to the
real numbers, we do not extended the execution modalities of
algorithms. We are still using the same execution modalities
---~Turing machines, rewrite systems, {\em etc.}  Simply, these
computation mechanisms are now used with rational numbers
approximating the real numbers.
The thesis that real computable functions defined in such a way
are sufficient to describe
the laws of nature, or that that an analog machine cannot compute
a function that would exceed this notion of computability, 
implies that all that an analog machine can compute
can also be computed by a digital machine, or that a physical process
cannot access in a finite time to the infinite amount of information
contained in a real number. This can be reformulated in more physical
terms as the fact that a physical process cannot be used to
distinguish, in a finite time, the difference between two
close enough magnitudes.

\subsection{The notion of a computable partial function over the real numbers}

This definition of computability over the real numbers 
generalizes easily to a notion of a computable partial function over the real
numbers. 

It suffices to modify the second condition according to which the sequence
$\varepsilon$ must go to $0$ at infinity.  Instead, we require that,
for each real number $x$, either for all sequences $(q_n)_n$,
$(\eta_n)_n$, $(r_n)_n$, $(\varepsilon_n)_n$ such that for all $n$,
$(r_n,\varepsilon_n) = F(q_n,\eta_n)$ and $|x - q_n| \leq \eta_n$ and
the sequence $\eta$ goes to $0$ at infinity, the sequence
$\varepsilon$ goes to $0$ at infinity, or for all sequences $(q_n)_n$,
$(\eta_n)_n$, $(r_n)_n$, $(\varepsilon_n)_n$ such that for all $n$,
$(r_n,\varepsilon_n) = F(q_n,\eta_n)$ and $|x - q_n| \leq \eta_n$ and
the sequence $\eta$ goes to $0$ at infinity, the sequence
$\varepsilon$ is constant and equal to $+ \infty$.  In this second
case, the attempt to obtain any approximation of $f(x)$ by providing
more and more accurate approximations of $x$ leads to a computation
that does not terminate: the function $f$ is not defined at $x$.

We can prove this way that the partial characteristic function of the
set ${\mathbb R}^{+} \setminus \{0\}$ is computable. Indeed, it is
computed by the 
function $F$ that maps $q$ and $\eta$ to the ordered pair $(1,\eta)$
if $q - \eta > 0$ and to the ordered pair $(1, + \infty)$ otherwise.
Indeed, if $x$ is a strictly positive real, $(q_n)_n$ and $(\eta_n)_n$
two sequences such that for all $n$, $|x - q_n| \leq \eta_n$ and
$\eta$ goes to $0$ at infinity, then the sequence $(q_n - \eta_n)_n$
goes to $x$ at infinity and thus it is strictly positive beyond a
certain point. The sequence $(F(q_n,\eta_n))_n$ is therefore equal to
$(1,\eta_n)_n$ beyond a certain point and $f(x) = 1$.  If $x$ is
negative or null, in contrast, we have for all $n$, $q_n - \eta_n \leq
x \leq 0$, thus the sequence $(q_n - \eta_n)_n$ is
always negative or null and the sequence $(F(q_n,\eta_n))_n$ is
always equal to $(1, + \infty)$: the function $f$ is not defined at
$x$. We can, this way, express formally what is often stated
informally: if $x$ is a strictly positive real, then there is an
effective way to prove it.

We can prove that the domain of a partial computable function is
always an open set. Indeed, if $f$ is a function and $x$ a real number
such that $f(x)$ is defined, then there exist sequences $(q_n)_n$,
$(\eta_n)_n$ such that for all $n$, $|x - q_n| \leq \eta_n$ and the
sequence $\eta$ goes to $0$ at infinity.  Let $(r_n)_n$ and
$(\varepsilon_n)_n$ be sequences such that for all $n$,
$(r_n,\varepsilon_n) = F(q_n,2 \eta_n)$. The function $f$ is defined
at $x$, the sequence $(2 \eta_n)_n$ verifies the condition that for
all $n$, $|x - q_n| \leq 2 \eta_n$ and it goes to $0$ at
infinity.  Thus, the sequence $\varepsilon$ goes to $0$ at
infinity and there exists a natural number $m$ such that
$\varepsilon_m$ is different from $+ \infty$. 
As $|x - q_m| \leq \eta_m$, 
the interval $[q_m - 2
\eta_m, q_m + 2 \eta_m]$ is a neighborhood of $x$ and for all $y$ in
this interval, $|y - q_m| \leq 2 \eta_m$, thus, there exist two
sequences $(q'_n)_n$ and $(\eta'_n)_n$ such that $q'_0 = q_m$ and
$\eta'_0 = 2 \eta_m$ and for all $n$, $|y - q'_n| \leq \eta'_n$ and
such that the sequence $\eta'$ goes to $0$ at infinity.  Let
$(r'_n)_n$ and $(\varepsilon'_n)_n$ be the sequences such that for all
$n$, $(r'_n,\varepsilon'_n) = F(q'_n,\eta'_n)$. We have
$\varepsilon'_0 = \varepsilon_m$ thus the sequence $\varepsilon'$ is
not constant and equal to $+ \infty$ and $f$ is defined at $y$.

Thus, if we call {\em semi-decidable} a set whose partial
characteristic function is computable, then all semi-decidable sets of
real numbers are open.

\section{Non deterministic algorithms over the real numbers}

We have extended above the notion of computability of functions operating on
natural numbers to relations between natural numbers.  We can try to
extend, in a similar way, the notion of computability of functions operating
on real numbers to relations between real numbers.

To guide us in the choice of an extension of the notion of computability 
of functions operating on real numbers to relations between real numbers, 
we shall require, as a minimal condition, that the
proposed definition extends the notion of a computable function, that is
that a functions operating on real numbers is computable as a function
if and only if it is computable as a relation.

\subsection{Representing the set $R_x$ by its characteristic function}

Representing the set $R_x$ by its characteristic function leads to
represent a relation $R$ by its characteristic function as well.

As computable functions over the real numbers are always continuous,
the relations that have a computable characteristic functions are only
the empty and the full relation.  Thus, this idea leads to dead end.

\subsection{Representing the set $R_x$ by its partial characteristic function}

Representing the set $R_x$ by its partial characteristic function leads to 
represent the relation $R$ by its partial characteristic function, as well.

We can, for instance, represent the relation $R$ such that 
$x~R~y$ if $x < y < x + 1$, whose partial characteristic function
is $\chi_{{\mathbb R}^+ \setminus \{0\}}((x + 1 - y)(y - x))$.  

In contrast, the partial characteristic function of the relation $R$
defined by $x~R~y$ if $y = x$ or $y = x + 1$ is not computable,
because its graph is not an open set.  In the same way, the partial
characteristic function of the functional relation $R$ defined by
$x~R~y$ if $y = x$ is not computable.
Thus, if this solution allows more relations to be represented than
the previous ones, it is far from being sufficient as the identity, that 
is computable as a function, would not be computable as a relation.

\subsection{Representing the set $R_x$ by an enumerating function}

As the characteristic function of equality over the real numbers is not
computable, the argument developed in the case of natural numbers that
effectively enumerable relations are semi-decidable does not
generalize to the case of real numbers: there are more
effectively enumerable relations than semi-decidable ones.

Representing the set $R_x$ by an enumerating function
leads to represent the relation $R$ by a function $f$ such that 
$x~R~y$ if and only if there exists an index $i$ such that 
$y = f(x,i)$. 
In this
definition, the set $\Omega$ may be arbitrary. It can, for instance,
be the set of natural numbers, the real line, an interval of the set
of natural numbers, an interval of the real line, {\em etc.}

This solution, unlike the previous ones, gives a large set of
representable relations, for instance the relation $R$ defined by
$x~R~y$ if $y = x$ or $y = x + 1$ is representable, taking the set of
natural numbers for the set $\Omega$ and, for $f$, the function
defined by $f(x,0) = x$ and $f(x,i) = x+1$ if $i \geq 1$.  This
function is computed by the function $F$ where $F(q,\eta, 0) =
(q,\eta)$ and $F(q,\eta, i) = (q+1,\eta)$ if $i \geq 1$.  The
functional relation $R$ defined by $x~R~y$ if $y = x$ can also be
represented by taking the set of natural numbers for $\Omega$ and, for
$f$, the function defined by $f(x,i) = x$, that is computed by the
function $F$ defined by $F(q,\eta, i) = (q, \eta)$.  

More generally, all computable functions $g$ mapping real numbers to
real numbers can be represented as relations, as it suffices to
define the function $f(x,i)$ as $g(x)$, ignoring the index
$i$. Conversely, a functional relation that can be represented is
computable as a function, because this function $g$ associates to $x$ the
value $f(x,i_0)$ where $i_0$ is an arbitrary index.

The conclusion of this section is this that among the three sets of
relations we have defined, the last one, the set of effectively
enumerable relations is the largest and it is the only one
that contains all computable functions and the relation that maps the
real number $x$ to the real numbers $x$ and $x + 1$, that intuitively seems
to be computable. It is the only one that can pretend to be used as a
description language for the laws of physics.

\section{Probabilities}

This notion of a non deterministic algorithm can be extended
to a notion of a probabilistic algorithm on real numbers.

Trying to represent a non deterministic algorithm as a computable
function $f$ mapping an ordered pair $(x,y)$ to the values $1$ of $0$
depending on the fact that $y$ could be the result of the algorithm
at $x$ or not, lead to failure because the function $f$ being
continuous, only the empty and full relations could be represented.

We can try, instead, to map the ordered pair $(x,y)$ not to $1$ or $0$
but to a real number indicating the propension of $y$ to be the result
of the algorithm at $x$.  Depending on the cases, this
propension can be a probability, a density of probability, 
a number whose square is such a probability or a density of
probability, {\em etc.}  

A probabilistic algorithm, that takes an element of a set $A$ as
argument and returns an element of a set $B$ if it terminates, defines
a function that maps every element of $A$ to a probability
distribution over the set $B$, that is a function that maps a pair
formed with an element $x$ of $A$ and a subset $Y$ of $B$ to the
probability that the value of this algorithm at $x$ is in $Y$.

Discussing the computability of this function is delicate, because, we
have the problem of representing the set $Y$.  We can nevertheless
focus on some particular cases.  If the set $Y$ is always an interval,
we can replace this function by a repartition function, that is a
function that maps $x$ and $y$ to the probability that the value of
the algorithm at $x$ is less than $y$.  If this repartition function
is moreover differentiable, we can replace it by its derivative, that
this the function that maps $x$ and $y$ to the density of probability
that the value of the algorithm at $x$ is $y$.  Another particular
case is when the algorithm can take only a finite or countable number
of values for each $x$. It is then possible to define this probability
distribution by a function that maps $x$ and $y$ to the probability
that the value of the algorithm at $x$ is exactly $y$.

If we require this function to be computable, it must be continuous.
In the first case, there are many random variables whose
repartition function or whose density function is continuous. In the
second, the fact that the function that maps $x$ and $y$ to the probability
that the value of the algorithm at $x$ is $y$ is continuous implies
that if it is different from zero at a point $y$, it must be minored 
by a strictly positive value on a neighborhood of $y$, which is
contradictory.  Thus, it is not possible to represent this way a
probabilistic algorithm that takes two different values at $0$ with
probabilities $1/2$.

Going from non deterministic algorithms to probabilistic ones thus
gives to this method based on characteristic functions a wider
spectrum, but not a spectrum wide enough to
describe algorithms with a discrete probability distribution.  The
same argument eliminates also the attempt to base a definition on
partial characteristic functions. A partial propension function
defined at a point $y$ is defined on a neighborhood of this point and
the same problem occurs as in the discrete case.

In contrast, the representation of relations based on the notion of
an effective enumeration generalizes easily to probabilistic algorithms 
with discrete probability distributions. 
To each pair $(x,i)$ we associate a pair $(y,p)$.
We can, for instance, represent the algorithm that maps the real
number $x$ to $x$ and $x + 1$ with probabilities $1/2$ by
taking for the set of indices $\Omega = \{0,1,2,3\}$
and by defining the function $f$ by 
$f(x,0) = f(x,1) = (x,1/4)$ and
$f(x,2) = f(x,3) = (x+1,1/4)$. 
The probability that the value of the
algorithm at $x$ is $y$ is the sum of the $p_i$ for the $i$ such that 
$y_i = y$. 
In this case, the probability that the result of the algorithm at $x$ 
is $x$ is $1/2$, that that it is $x + 1$ is $1/2$ as well, and 
that that it is another value is $0$. 
This function that takes its value in a Cartesian product can be
decomposed into two functions $h$ and $p$ such that $h(x,0) = h(x,1) =
x$, $h(x,2) = h(x,3) = x + 1$ and $p(x,0) = p(x,1) = p(x,2) = p(x,3) =
1/4$.  This amounts to define a non deterministic algorithm with the
function $h$ and to equip independently the set $\Omega$ with a
probability distribution.

This solution also permits to represent the algorithm that maps the
real number $x$ to $x$ and $2x$ with 
probabilities $1/2$ and the real number $0$ to $0$ with
probability $1$, defining the function $f$ by $f(x,0) = f(x,1) =
(x,1/4)$ and $f(x,2) = f(x,3) = (2x,1/4)$. This would not be allowed 
with a solution that would represent this algorithm as a function
mapping each $x$ to the set of possible results, each with its
probability, as it would be discontinuous at $0$.

The conclusion of this section is thus that, in the case of
probabilistic algorithms, like in the case of non deterministic
algorithms, the definition based on effective enumeration is that that
permits to represent the largest set of relations. In particular, it
is the only one that permits to represent the relation that maps the
real number $x$ to $x$ and $x + 1$, with 
probabilities $1/2$, that intuitively seems to be computable.

\section{What is a computable description of a non deterministic
theory in physics?}

This discussion on the nature of non deterministic algorithms on
real numbers permits to describe more precisely what it could mean for a
theory in physics to have an algorithmic expression or not, when this
theory is non deterministic and represents physical magnitudes with real
numbers.

To do so, let us imagine an experiment in which one prepares a system
by choosing a magnitude $x$, let the system evolve for a fixed time
$T$, and measures a magnitude $y$.  Our two hypotheses on the theory
we try to describe are that it represents magnitudes $x$ and $y$ by
real numbers and that for each value $x$, it does not prescribe a
unique result $y$, but a set of possible results, possibly
equipped with probabilities.  What would be an algorithm that would
describe the results prescribed by the theory, for such an experiment?

As we have seen, we should expect such an algorithm to indicate
neither if $y$ is a possible result for the experiment initiated with
the value $x$, nor the probability for this result to be $y$.

In contrast, the theory should describe this experiment by a function
that to each value of $x$ associates a random variable 
$f_x$ that is itself a function that maps each element of a set of
indices $\Omega$ to a value $y$. We can call $f$ the function 
that maps  $x$ and $i$ to the value $f_x(i)$.  It is this function, 
that an algorithmic description of the theory 
must represent by an algorithm.

The existence of such an algorithm is a strong constraint on the
theory. In all the cases where the experiment is deterministic, it
boils down exactly to the deterministic physical Church-Turing thesis: the
fact that the link between $x$ and $y$ is a computable function.  It
is therefore neither more constraining nor less constraining for a
theory to be algorithmic in the non deterministic case than in the
deterministic case.

We shall not answer here to the question of whether such or such
theory in physics is algorithmic or not. Our only purpose in this
paper was to try to understand how this question could be stated.

\section*{Acknowledgments}

I want to thank Pablo Arrighi and Jean-Baptiste Joinet for many
remarks on a previous draft of this paper, Olivier Bournez, Assia
Mahboubi and Nathalie Revol for helping me to find my way in the
complex domain of real computation and Giuseppe Longo and Thierry Paul
for always stimulating discussions.

\end{document}